\documentclass[final]{svjour3}                     
\usepackage[numbers]{natbib}
\usepackage{amsmath,amsfonts} 
\usepackage{amssymb}
\usepackage{array}
\usepackage[tight,TABTOPCAP]{subfigure}
\usepackage{fixltx2e} 
\usepackage{graphicx}
\newcommand{\degree}{\ensuremath{^\circ}}

\newcommand{\be}{\begin{equation}}
\newcommand{\ee}{\end{equation}}
\newcommand{\ba}{\begin{eqnarray}}
\newcommand{\ea}{\end{eqnarray}}
\newcommand{\ban}{\begin{eqnarray*}}
\newcommand{\ean}{\end{eqnarray*}}

\usepackage{tikz}
\tikzset{
  pics/.cd,
  disc/.style = { 
    code = {
      \path [top color = black!25, bottom color = white] 
        (0,.005) ellipse [x radius = 0.9, y radius = 0.6];
      \path [left color = black!25, right color = black!25, middle color = white] 
        (-0.9,0) -- (-0.9,-2) arc (180:360:0.9 and 0.6)
               -- (0.9,0) arc (360:180:0.9 and 0.6);
      \foreach \r in {225,315}
        \foreach \i [evaluate = {\s=30;}] in {0,2,...,30}
          \fill [black, fill opacity = 1/50] 
            (0,0.02) -- (\r+\s-\i:0.9 and 0.6) -- ++(0,-2) 
            arc        (\r+\s-\i:\r-\s+\i:0.9 and 0.6) -- ++(0,2) -- cycle;
      \foreach \r in {45,135}
        \foreach \i [evaluate = {\s=30;}] in {0,2,...,30}
          \fill [black, fill opacity = 1/50] 
            (0,0.02) -- (\r+\s-\i:0.9 and 0.6) 
            arc (\r+\s-\i:\r-\s+\i:0.9 and 0.6)  -- cycle;
    }
  },
    rc/.style = {
    code = {
      \path [top color = black!25, bottom color = white] 
        (0,.005) ellipse [x radius = 0.5, y radius = 0.34];
      \path [left color = black!25, right color = black!25, middle color = white] 
        (-0.5,0) -- (-0.5,-2) arc (180:360:0.5 and 0.34)
               -- (0.5,0) arc (360:180:0.5 and 0.34);
      \foreach \r in {225,315}
        \foreach \i [evaluate = {\s=30;}] in {0,2,...,30}
          \fill [black, fill opacity = 1/10] 
            (0,0.2) -- (\r+\s-\i:0.5 and 0.34) -- ++(0,-2) 
            arc        (\r+\s-\i:\r-\s+\i:0.5 and 0.34) -- ++(0,2) -- cycle;
      \foreach \r in {45,135}
        \foreach \i [evaluate = {\s=30;}] in {0,2,...,30}
          \fill [black, fill opacity = 1/10] 
            (0,0.2) -- (\r+\s-\i:0.5 and 0.34) 
            arc (\r+\s-\i:\r-\s+\i:0.5 and 0.34)  -- cycle;
    }
  },
    shell/.style = { 
    code = {
      \path [top color = black!25, bottom color = white] 
        (0,.05*6/3) ellipse [x radius = 3-.05, y radius = 2-.032];
      \path [left color = black!25, right color = black!25, middle color = white] 
        (-3,0) -- (-3,-2) arc (180:360:3 and 2)
               -- (3,0) arc (360:180:3 and 2);
      \foreach \r in {225,315}
        \foreach \i [evaluate = {\s=30;}] in {0,2,...,30}
          \fill [black, fill opacity = 1/50] 
            (0,0.2) -- (\r+\s-\i:3 and 2) -- ++(0,-2) 
            arc        (\r+\s-\i:\r-\s+\i:3 and 2) -- ++(0,2) -- cycle;
      \foreach \r in {45,135}
        \foreach \i [evaluate = {\s=30;}] in {0,2,...,30}
          \fill [black, fill opacity = 1/50] 
            (0,0.2) -- (\r+\s-\i:3 and 2) 
            arc (\r+\s-\i:\r-\s+\i:3 and 2)  -- cycle;
      \path [top color = white, bottom color = white] 
        (0,.05*6/3) ellipse [x radius = 2-.05, y radius = 4/3-.05*4/3];
    }
  },
    cylinder/.style = { 
    code = {
      \path [top color = black!25, bottom color = white] 
        (0,.05*6/3) ellipse [x radius = 3-.05, y radius = 2-.032];
      \path [left color = black!25, right color = black!25, middle color = white] 
        (-3,0) -- (-3,-2) arc (180:360:3 and 2)
               -- (3,0) arc (360:180:3 and 2);
      \foreach \r in {225,315}
        \foreach \i [evaluate = {\s=30;}] in {0,2,...,30}
          \fill [black, fill opacity = 1/50] 
            (0,0.2) -- (\r+\s-\i:3 and 2) -- ++(0,-2) 
            arc        (\r+\s-\i:\r-\s+\i:3 and 2) -- ++(0,2) -- cycle;
      \foreach \r in {45,135}
        \foreach \i [evaluate = {\s=30;}] in {0,2,...,30}
          \fill [black, fill opacity = 1/50] 
            (0,0.2) -- (\r+\s-\i:3 and 2) 
            arc (\r+\s-\i:\r-\s+\i:3 and 2)  -- cycle;
    }
  }
}

\begin{document}

\title{Dual-band Dielectric Light-harvesting Nanoantennae \\ Made by Nature}
\author{Julian Juhi-Lian TING}
\institute{De-Font Research Institute, Taichung 40344, Taiwan, R.O.C. \\\email {juhilian@gmail.com \;\;\;\;URL:~http://amazon.com/author/julianting}}
\dedication{to the memory of my father.}
\maketitle

\begin{abstract}
Mechanisms to use nanoparticles to separate sunlight into photovoltaic useful range and thermally useful range to increase the efficiency of solar cells
and to dissipate heat radiatively are discussed based upon lessons we learnt from photosynthesis.
We show that the dual-band maxima in the absorption spectrum of bacterial light harvestors not only are due to the 
bacteriochlorophylls involved but also come from the geometry of the light harvestor.
Being able to manipulate these two bands arbitrarily enables us to fabricate the nanoparticles required.
Such mechanisms are also useful for the design of remote power charging and light sensors.



\end{abstract}

\newpage

\section{Introduction}

Waste heat disposal is an important issue in everyday life, ranging from our personal computers to power plants.
Recently, two seemingly unrelated studies in this topic actually employed the same mechanism which our previous study on bacteria light harvestors can provide
theoretical guidelines.
The first one concerns about the conversion efficiency of solar cell, while the second one is a general waste heat dissipation method~\citep{Hjerrild2016,Zhai2017}.

Solar-cell production is a mature industry today.
Trying to improve the efficiency of solar cells with the material used is as difficult as trying to raise the critical temperature of 
high-temperature superconductors, not even to mention there is a Shockley-Queisser limit~\citep{Shockley1961}.
However, the conversion efficiency of solar cells will be reduced 
by 0.4-0.5 \% for each one degree Celsius raised above their normal working temperature, i.e., $25 \degree C$~\citep{Wysocki1960,King1997,Dupre2015},
which is unavoidable because some frequencies of light are not useful for the solar cell and will be converted into heat eventually.
Various methods trying to reduce the working temperature have been proposed~\citep{Oh2018}.
An ingenious method to pre-filter the sunlight involves spreading of nanoparticles into water to separate the light into a
photovoltaically useful range and a thermally useful range has been proposed~\citep{Hjerrild2016}.
A similar general cooling method uses the reversed mechanism, which embeds particles into polymethylpentene films to enhance the radiation,
was also proposed~\citep{Zhai2017}.
Although these authors arrived at their proposal mainly through experimental trial and error, 
there exist theories to design what kind of particles should be used to achieve the best performance~\citep{Ting2018,Jain2018}.

One can readily recognise that the particles used by the latter authors are antennae, 
but what the former authors called filters are also better described as antennae, in particular, nanoantennae, 
which are devices designed to convert free radiation efficiently into localised energy, such as with lenses and mirrors, or vice versa~\citep{Novotny2017}.
The theory of antennae sums up individual effects of the light-matter interaction, 
in which the most important parameter is the geometry that produces the retarded effects.
Receiving nanoantennae may confine incident light into subwavelength region.
Optical antennae pose special problems because of their small size and the plasmon resonances of metal.
A consideration of nanoantennae as solar energy collectors has appeared previously~\cite{Kotter2010}.

Many effective light absorbing molecules are made in nature.
Chlorophylls interact with light stronger than others because they consist of alternating chemical double bonds and densely
packed to enhance the absorption cross section.
The structures of many bacteria light-harvesting complexes (LH) made of bacteriochlorophylls have been solved 
after McDermott {\it et al.} solved the structure of {\it Rhodopseudomonas acidophlla} in 1995~\citep{McDermott1995}.
Two of them are shown in the Table \ref{more}.

Understanding the light harvestor made by nature can not only
help us to fabricate voltaic solar cells 
but can also help us to manufacture optical wireless power chargers~\citep{Elsheakh2017} and make better electromagnetic wave sensors~\citep{Watanabe2015}.
Wireless power transfers, at this moment, work not only under microwave frequencies, if not radio frequencies, 
but also only under designed non-physiological condition~\citep{Shinohara2014,Miller2014,Keskin2015,Tsai2016}.
However, medical applications require the power chargers to work under flexible physiological condition~\citep{Miller2014}.
The next generation wireless power transfer we discuss, will use incoherent light as the nature use, which is sustainable and renewable,
and is robust and, in particular, is working under physiological condition.

In two previous papers, the non-reciprocal properties and the dipole properties of the bacteria light harvestors, LH1-RC/LH2 complexes,
were considered by an infinitesimally thin loop antenna as a first approximation~\citep{Ting2018b,Ting2018}.
Even though that loop antenna considered is too idealized, it can explain many phenomena and natural design physically
without {\it ad hoc} parameter.
However, even though the resonance frequency can be adjusted by various methods~\cite{Alu2008}, 
a loop-antenna has only one resonance frequency, 
whereas a real light-harvesting complex shows not only a band of absorption frequency 
but a dual-band of absorption spectrum, 
with one in a region about $850-860 nm$ and the other near $800 nm$, 
which are generally attributed to the resonance frequencies of the bacteriochlorophylls involved~\citep{Wu1997,Georgakopoulou2002}.
There are indeed many methods to make a dual-band antenna~\citep{Pathak2017}, but the simplest way can be employed by the nature seems to come from
the geometry of the light harvestor.
We will show that there are geometrical reasons for these two maxima by extending our analysis
to consider the finite thickness of the antenna.
Being able to manipulate these two bands arbitrarily enable us to fabricate the nanoparticles required.

To simplify the issue, we will consider a LH2 system, without the feeding line and the reaction centre of the LH1 system, in the following.
A LH2 consists of nine units for 
{\it Rhodopseudomonas acidophila} of inner diameter $36$ \AA ~and outer diameter $68$ \AA ~\citep{McDermott1995,Papiz2003}.
The width is roughly double the height of the molecule.
The dimensions of the molecule are obtainable 
using the data from the Protein Data Bank (PDB, http://www.rcsb.org/pdb/).
Such systems are exactly what Hjerrild {\it et al.} and Zhai {\it et al.} considered, 
except that nature made the light harvestors with a few atoms whereas
the following discussion considers the particles with bulk material as the previous authors.
Because the light harvestors have random orientation, it would be better to consider their interaction with unpolarized light.
The energy received by such antennae will be dissipated as heat into surrounding water.

\begin{table}[tb]
\centering
\begin{tabular}{p{0.6\columnwidth}lcp{0.15\columnwidth}}
\hline 
protein & PDB ID & symmetry & cartoon \\
\hline \\



LH2 B800-850 from Rhodopseudomonas acidophila 
\citep{Papiz2003}              & 1NKZ & C9 &
\begin{minipage}[c]{0.15\columnwidth}\includegraphics[width=\textwidth, angle=0]{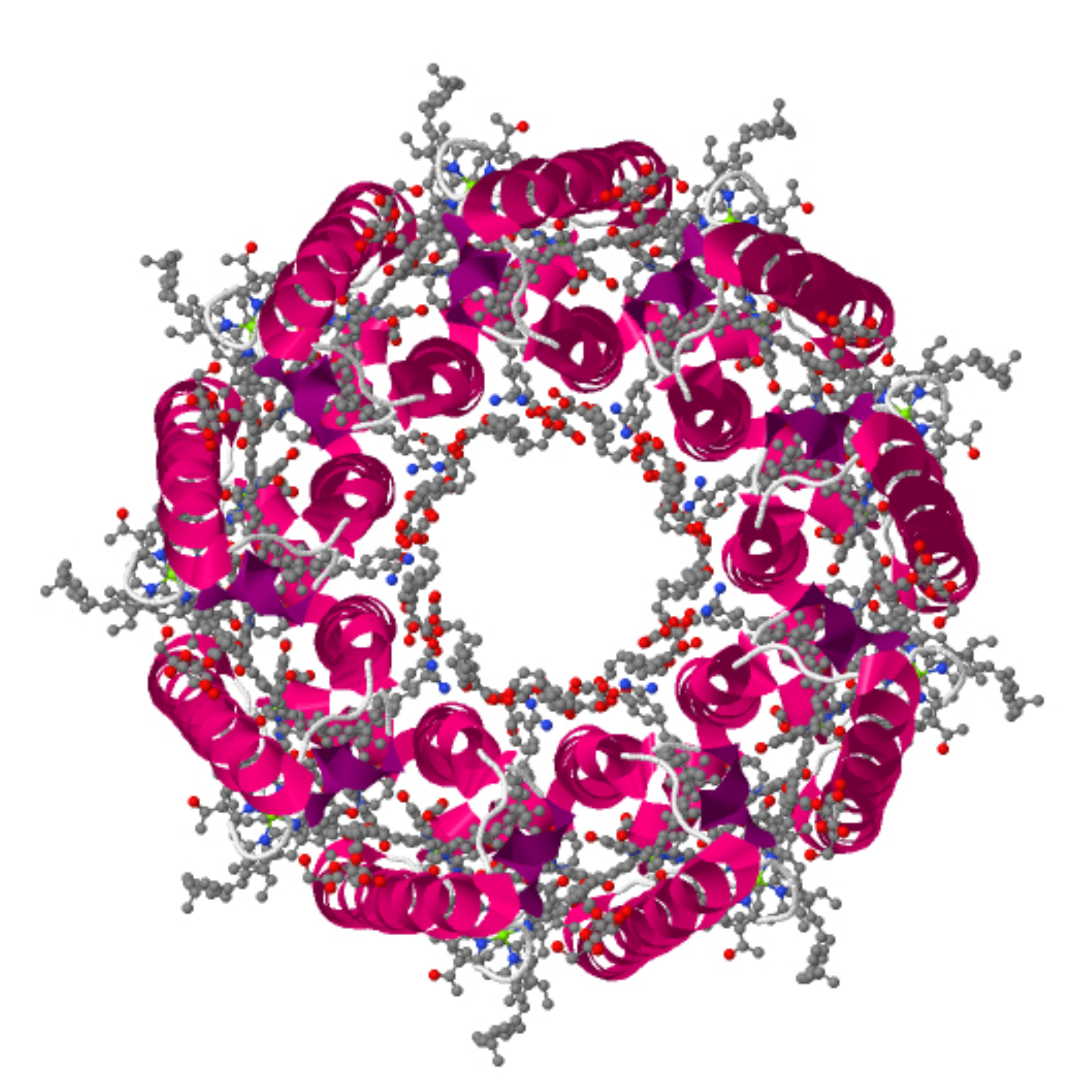}\end{minipage}  \\

LH2 B800-850 from Rhodospirillum molischianum \citep{Koepke1996}              & 1LGH & C8 &

\begin{minipage}[c]{0.15\columnwidth}\includegraphics[width=\textwidth, angle=0]{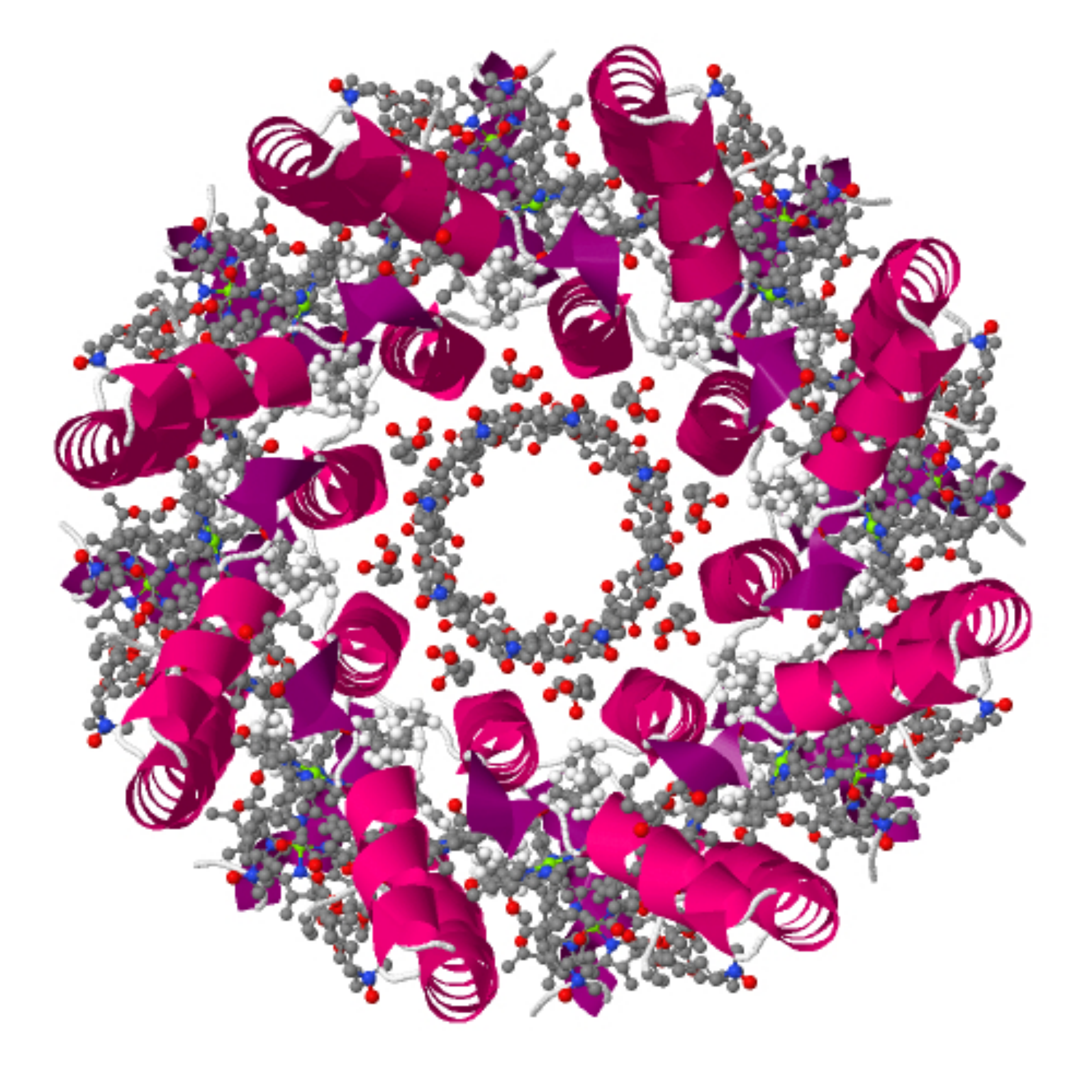}\end{minipage}  \\

\hline 
\end{tabular}
\caption{Various light harvestors II with structural symmetry. More can be found in PDB.
}
\label{more}
\end{table}

\section{dielectric light harvestor}
\label{dielectric}

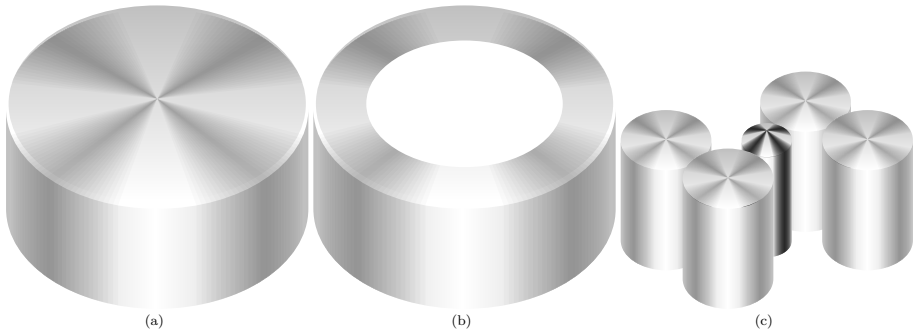
\begin{figure}[tb]
\begin{center}

\resizebox{\columnwidth}{!}{
\mbox{
\subfigure[]{\begin{tikzpicture}\path (0,0,0) pic {cylinder} ; \end{tikzpicture}}
\subfigure[]{\begin{tikzpicture}\path (0,0,0) pic {shell} ; \end{tikzpicture}}
\subfigure[]{\begin{tikzpicture}\path (0,0,0) pic {disc}  (2,0,-2) pic {disc} (2,0,0) pic {rc} (4,0,0) pic {disc} (2,0,2) pic {disc} ;\end{tikzpicture}}
}
}
\caption{Successive approximation of the bacterial light-harvesting antenna: 
(a) cylindrical approximation,
(b) dielectric shell (ring) approximating an antenna, 
(c) four-element dielectric antenna with centre feed line. 
}
\label{cylinder}
\end{center}
\end{figure}

Optical nanoantennae have been constructed in 2005~\citep{Schuck2005,Farahani2005}.
Albeit most nanoantennae are as yet made of conductors~\citep{Bharadwaj2009,Monticone2017},
at the frequency of solar light, metals are no longer perfect conductors hence the electric fields can
penetrate inside, and dielectrics work better. 
Paraphrasingly, the conductivity turns out to be less relevant when the wavelength is reduced.
Dielectric antennae are at present widely used in microwave engineering, 
optical-fibre engineering and, in particular, in mobile telephones~\citep{Yoon2010,Lee2014,Monticone2017}.
A dielectric antenna functions like an acoustic resonator; the electrons bounce back and forth inside the cavity, 
which is consistent with the exciton theories.
All-dielectric nanoantennae have been fabricated recently~\citep{Li2015}.

Presently people use conductor coated with dielectric to make such antenna, which is unnecessary.
Using a dielectric material to build the light harvestor has many advantages~\citep{Mongia1994,Kuznetsov2016}:
\begin{itemize}
\item 

the dimension of the antenna can be diminished by a factor
$1/\sqrt{\epsilon_r}$ relative to the corresponding metallic one because
the dielectric wave length is smaller than the free-space wavelength,

\item relatively large bandwidth,
\item low or no loss at optical frequencies,
\item 
free from damage of high power or local heating~\citep{Hsu2007a},

\item 
have less detuning when placed close to another object,

\item compatibility with complementary metal-oxide semiconductor fabrication processes.
\end{itemize}
Symmetry breaking of the electric field in space facilitates the radiation of the antenna~\citep{Sinha2015}, which
provides a third reason for the mysterious opening at the LH1-RC complex and explains the elliptical shape of the complex.

We made successive approximations for the light harvestor in Figure \ref{cylinder}.
The first antenna, Figure \ref{cylinder} (a), is a cylinder, which can be solved analytically,
but the boundary condition that we consider differs slightly from standard engineering practice 
because in most antenna applications the dielectric material is placed on top of a magnetic conducting plate,
whereas in photosynthesis the dielectric material is placed in a solvent.
This does not cause much difficulty to apply the solutions found in the literature. 
Using the method of image, we simply consider a cylinder of double-height,
$2h$, so that the solution for an antenna placed on top of a magnetic conducting plate can be applied.
The second one, a dielectric ring (or shell) antenna, as shown in Figure \ref{cylinder}(b), can also be solved analytically with a superposition principle.
The third one shown in Figure \ref{cylinder} (c) generally requires numerical methods of solution~\citep{Guha2006}.

We consider first a solid cylindrical dielectric material of diameter $a$ extending in direction $z$ from $-h$ to $h$ in a vacuum.
Fields at the surface of a region of large relative permittivity satisfy approximately
an open-circuit boundary condition, i.e. the normal component of the electric field and the tangential component of the magnetic
field are zero~\citep{Ramo1994}.
The resonance frequencies of least order read~\citep{Mongia1994}:
{\small
\ban
f_0 (HEM_{01}) &\approx & \dfrac{c 6.323}{2 \pi a \sqrt{\epsilon_r +2}} \left[ 0.27 + 0.36 \left(\dfrac{a}{2h}\right) + 0.02 \left(\dfrac{a}{2h}\right)^2 \right]\\
f_0 (TE_{01\delta}) &\approx & \dfrac{c 2.327}{2 \pi a \sqrt{\epsilon_r +1}} \left[ 1 + 0.2123 \left(\dfrac{a}{h}\right) - 0.00898 \left(\dfrac{a}{h}\right)^2 \right]\\ 
f_0 (TM_{01\delta}) &\approx & \dfrac{c }{2 \pi a \sqrt{\epsilon_r +2}} \left[ 3.83^2 + \left(\dfrac{\pi a}{2h}\right)^2 \right]^{1/2}\\
f_0 (TE_{011+\delta}) &\approx & \dfrac{c 2.280}{2 \pi a \sqrt{\epsilon_r +1}} \left[ 1 + 0.7013 \left(\dfrac{a}{h}\right)-0.002713 \left(\dfrac{a}{h}\right)^2 
 \right]^{1/2}
\ean
}
with $\epsilon_r = \epsilon/\epsilon_0$ been the relative permittivity, and $c$ is the speed of light.
As we stated at the beginning, our $a/h \approx 2$ is well within the valid range of the approximation, which controls the bandwidth of the antenna\citep{DeYoung2006}.
Depends upon the relative permittivity, at a certain ratio of $a/h$ the bandwidth increased due to overlapping of two resonant modes.
We use $h=34 $\AA, $a = 68$ \AA ~and $\epsilon_r =1000$ to obtain the values shown in the first row of Table \ref{freq},
but this result is not yet what we want. 
What we seek is a shell (or ring) shape, which can be effected through superposition of two cylinders of different radius~\citep{Guo2005}.
The modes of a smaller cylindrical dielectric  material of diameter $b$ with the same height $2h$ can be obtained similarly and are shown in the second row of Table \ref{freq}.
The superposition produces an antenna with two maxima in the absorption spectrum, which is called
a dual-band antenna, as the bacteria light harvestors also have two maxima in their absorption spectrum~\citep{Law2004}.
These numbers are roughly what we require, as the relative permittivity for the light harvestor do not exist~\citep{Alden1997}.

The relative permittivity appears to be spurious high, 
because we are using radio frequency formulae for bulk material properties.
At optical frequencies even matal is no longer perfect conductors.
The skin depth $d= \lambda /(4 \pi \sqrt{\epsilon_r}) \approx 20 ~\AA$.
Electricmagnetic fields have fully penetrated the antenna of thickness $30 ~ \AA$.
The volume currents and reduced wavelength for optical antennae need to be adjusted by about 20\% ~\citep{Novotny2007}.
Furthermore, the nature is not using bulk material.
However, even without these adjustments we have already shown it is possible to produce the dual-band spectrum.
A similar circular polarized antenna excited with a tilted modified square slot has been studied~\citep{Pathak2017}.

\begin{table}[tb]
\centering
\begin{tabular}{lcccc}
\hline 
radius/mode        &  $TE_{01\delta}$  & $ HEM_{11\delta}$ & $ TM_{01\delta}$ & $  TE_{011+\delta} $ \\
\hline 
$68~ $\AA           &  0.717            & 0.912             &  1.099           & 1.172      \\
$36~ $\AA           &  1.185            & 1.235             &  1.75            & 1.61       \\
\hline
\end{tabular}
\caption{Various mode resonance frequencies/ GHz with $h=34 $\AA ~and $\epsilon_r =1000$.
}
\label{freq}
\end{table}

For two reasons it would be better to divide the light harvestor into sub-units as shown in Figure \ref{cylinder} (c):
\begin{itemize}
\item For a dielectric antenna to achieve a satisfactory efficiency, 
proper modes must be excited by sunlight, as shown in Figure 5 of Yu {\it et al.}~\citep{Yu2012}.
Modules of bacteriochlorophyll appear at nodes of the electric field~\citep{Yu2012,Soren2014}.
/
\item

A ring that is constructed from sub-units corresponds to an interconnection of inductors interleaved by capacitors works better than a continuous one, 
because it guides the flow of displacement current better~\citep{Alu2008a}.
Furthermore, it has a band-stop frequency response that rejects a certain range of frequency.

\end{itemize}

A traditional theory of photosynthesis describes this part according to exciton theory without detail.
In view of this second reason, the light harvestor built by previous authors apparently has room for improving their efficiency~\citep{Yu2012,Hjerrild2016,Hjerrild2017}.

\section{Summary}

Historically, many authors tried to calculate spectroscopic properties from some theoretical models as the spectrum is
readily accessible experimentally~\citep{Alden1997}.
In this work, we show 
that it is possible to place two maxima of the spectrum at the position desired by the classical electrodynamics model we proposed.
An important difference between radiowave and optical antenna is at the way the receiver is connected to the antenna
as shown by the LH1 designed by nature.
We can also have a multi-element multi-layer model, which resembles the light harvestor even better, with
radiation or receiving pattern achieving a monopole-like efficiency~\citep{Chaudhary2011}.
Soren {et al.} considered a Sierpinski carpet patterned cylindrical Dielectric Resonator Antenna made by Teflon, which resemble our
light-harvesting antennae even better~\citep{Soren2012}.

The nanoparticles that Hjerrild {\it et al.} considered are disk-shaped metallic
particles of diameter roughly $100 nm$, which exploit plasmon resonance. 
We show that the optimal shape of the particles should be toroidal instead of discotic or nanorods for the infrared range~\citep{Ting2018,Ting2018b}.
These authors further pointed out that silver is the best material to use, but that is quite expensive.
They considered the material used to be important for the frequencies of absorption, and mentioned that the ratio
of surface to volume of these nanoparticles makes the particles susceptible to damage from high power.
Therefore they coated their metallic disk with dielectric.
We have shown that full dielectric antennae can be used to avoid such high-power damage, 
which is actually the best material at such a scale, and that the frequency of absorption can be tailored with the radii of the toroidal particles.


\bibliography{d:/library} 

\begin{thebibliography}{10}
\providecommand{\url}[1]{{#1}}
\providecommand{\urlprefix}{URL }
\expandafter\ifx\csname urlstyle\endcsname\relax
  \providecommand{\doi}[1]{DOI \discretionary{}{}{}#1}\else
  \providecommand{\doi}{DOI \discretionary{}{}{}\begingroup
  \urlstyle{rm}\Url}\fi

\bibitem{Hjerrild2016}
N.E. Hjerrild, S.~Mesgari, F.~Crisostomo, J.A. Scott, R.~Amal, R.A. Taylor,
  Sol. Energy Mater. Sol. Cells \textbf{147}, 281 (2016).
\newblock \doi{10.1016/j.solmat.2015.12.010}

\bibitem{Zhai2017}
Y.~Zhai, Y.~Ma, S.N. David, D.~Zhao, R.~Lou, G.~Tan, R.~Yang, X.~Yin, Science
  (80-. ). \textbf{355}(6329), 1062 (2017).
\newblock \doi{10.1126/science.aai7899}

\bibitem{Shockley1961}
W.~Shockley, H.J. Queisser, J. Appl. Phys. \textbf{32}(3), 510 (1961).
\newblock \doi{10.1063/1.1736034}

\bibitem{Wysocki1960}
J.J. Wysocki, P.~Rappaport, J. Appl. Phys. \textbf{31}(3), 571 (1960).
\newblock \doi{10.1063/1.1735630}

\bibitem{King1997}
D.~King, J.~Kratochvil, W.~Boyson, in \emph{Conf. Rec. Twenty Sixth IEEE
  Photovolt. Spec. Conf. - 1997} (IEEE, 1997), pp. 1183--1186.
\newblock \doi{10.1109/PVSC.1997.654300}

\bibitem{Dupre2015}
O.~Dupr{\'{e}}, R.~Vaillon, M.~Green, Sol. Energy Mater. Sol. Cells
  \textbf{140}, 92 (2015).
\newblock \doi{10.1016/j.solmat.2015.03.025}

\bibitem{Oh2018}
J.~Oh, B.~Rammohan, A.~Pavgi, S.~Tatapudi, G.~Tamizhmani, G.~Kelly, M.~Bolen,
  IEEE J. Photovoltaics \textbf{8}(5), 1160 (2018).
\newblock \doi{10.1109/JPHOTOV.2018.2841511}

\bibitem{Ting2018}
J.J.L. Ting, J. Photochem. Photobiol. B Biol. \textbf{179}, 134 (2018).
\newblock \doi{10.1016/j.jphotobiol.2018.01.011}

\bibitem{Jain2018}
P.K. Jain, Phys. Today \textbf{71}(8), 10 (2018).
\newblock \doi{10.1063/PT.3.3984}

\bibitem{Novotny2017}
L.~Novotny, B.~Hecht, \emph{{Principles of Nano-Optics}} (Cambridge University
  Press, Cambridge, 2006).
\newblock \doi{10.1017/CBO9780511813535}

\bibitem{Kotter2010}
D.K. Kotter, S.D. Novack, W.D. Slafer, P.J. Pinhero, J. Sol. Energy Eng.
  \textbf{132}(1), 011014 (2010).
\newblock \doi{10.1115/1.4000577}

\bibitem{McDermott1995}
G.~McDermott, S.M. Prince, A.A. Freer, A.M. Hawthornthwaite-Lawless, M.Z.
  Papiz, R.J. Cogdell, N.W. Isaacs, Nature \textbf{374}(6522), 517 (1995).
\newblock \doi{10.1038/374517a0}

\bibitem{Elsheakh2017}
D.~Elsheakh, in \emph{Microw. Syst. Appl.}, ed. by S.K. Goudos (IntechOpen,
  Rijeka, 2017), chap.~08, pp. 155--205.
\newblock \doi{10.5772/64918}

\bibitem{Watanabe2015}
M.~Watanabe, A.~Nakamura, A.~Kunii, K.~Kusano, M.~Futagawa, J. Phys. Conf. Ser.
  \textbf{660}(1), 0 (2015).
\newblock \doi{10.1088/1742-6596/660/1/012110}

\bibitem{Shinohara2014}
N.~Shinohara, \emph{{Wireless Power Transfer via Radiowaves}}.
\newblock ISTE (John Wiley {\&} Sons, Inc., Hoboken, NJ, USA, 2013).
\newblock \doi{10.1002/9781118863008}

\bibitem{Miller2014}
J.L. Miller, Phys. Today \textbf{67}(8), 12 (2014).
\newblock \doi{10.1063/PT.3.2464}

\bibitem{Keskin2015}
N.~Keskin, H.~Liu, in \emph{2015 IEEE 65th Electron. Components Technol. Conf.}
  (IEEE, 2015), Dc, pp. 1828--1833.
\newblock \doi{10.1109/ECTC.2015.7159848}

\bibitem{Tsai2016}
J.S. Tsai, J.S. Hu, S.L. Chen, X.~Huang, Adv. Mech. Eng. \textbf{8}(2), 1
  (2016).
\newblock \doi{10.1177/1687814016632693}

\bibitem{Ting2018b}
J.J.L. Ting,  p. 1702.06671 (2019)

\bibitem{Alu2008}
A.~Al{\`{u}}, N.~Engheta, Phys. Rev. Lett. \textbf{101}(4), 043901 (2008).
\newblock \doi{10.1103/PhysRevLett.101.043901}

\bibitem{Wu1997}
H.M. Wu, N.R.S. Reddy, G.J. Small, J. Phys. Chem. B \textbf{101}(4), 651
  (1997).
\newblock \doi{10.1021/jp962766k}

\bibitem{Georgakopoulou2002}
S.~Georgakopoulou, R.N. Frese, E.~Johnson, C.~Koolhaas, R.J. Cogdell, R.~van
  Grondelle, G.~van~der Zwan, Biophys. J. \textbf{82}(4), 2184 (2002).
\newblock \doi{10.1016/S0006-3495(02)75565-3}

\bibitem{Pathak2017}
D.~Pathak, S.K. Sharma, V.S. Kushwah, Prog. Electromagn. Res. M
  \textbf{62}(November), 123 (2017).
\newblock \doi{10.2528/PIERM17092701}

\bibitem{Papiz2003}
M.Z. Papiz, S.M. Prince, T.~Howard, R.J. Cogdell, N.W. Isaacs, J. Mol. Biol.
  \textbf{326}(5), 1523 (2003).
\newblock \doi{10.1016/S0022-2836(03)00024-X}

\bibitem{Koepke1996}
J.~Koepke, X.~Hu, C.~Muenke, K.~Schulten, H.~Michel, Structure \textbf{4}(5),
  581 (1996).
\newblock \doi{10.1016/S0969-2126(96)00063-9}

\bibitem{Schuck2005}
P.J. Schuck, D.P. Fromm, A.~Sundaramurthy, G.S. Kino, W.E. Moerner, Phys. Rev.
  Lett. \textbf{94}(1), 017402 (2005).
\newblock \doi{10.1103/PhysRevLett.94.017402}

\bibitem{Farahani2005}
J.N. Farahani, D.W. Pohl, H.J. Eisler, B.~Hecht, Phys. Rev. Lett.
  \textbf{95}(1), 17402 (2005).
\newblock \doi{10.1103/PhysRevLett.95.017402}

\bibitem{Bharadwaj2009}
P.~Bharadwaj, B.~Deutsch, L.~Novotny, Adv. Opt. Photonics \textbf{1}(3), 438
  (2009).
\newblock \doi{10.1364/AOP.1.000438}

\bibitem{Monticone2017}
F.~Monticone, C.~Argyropoulos, A.~Alu, IEEE Antennas Propag. Mag.
  \textbf{PP}(99), 1 (2017).
\newblock \doi{10.1109/MAP.2017.2752721}

\bibitem{Yoon2010}
S.~Yoon, C.~Park, M.~Kim, K.~Kim, Y.~Yang, in \emph{2010 Asia-Pacific Microw.
  Conf.} (2010), pp. 219--222

\bibitem{Lee2014}
J.~Lee, J.~Lee, K.~Min, Y.~Cheon, IEEE Antennas Wirel. Propag. Lett.
  \textbf{13}, 935 (2014).
\newblock \doi{10.1109/LAWP.2014.2323066}

\bibitem{Li2015}
Z.~Li, X.~Liu, N.~Xu, J.~Du, Phys. Rev. Lett. \textbf{114}(14), 1 (2015).
\newblock \doi{10.1103/PhysRevLett.114.140504}

\bibitem{Mongia1994}
R.K. Mongia, P.~Bhartia, Int. J. Microw. Millimeter-Wave Comput. Eng.
  \textbf{4}(3), 230 (1994).
\newblock \doi{10.1002/mmce.4570040304}

\bibitem{Kuznetsov2016}
A.I. Kuznetsov, A.E. Miroshnichenko, M.L. Brongersma, Y.S. Kivshar,
  B.~Luk'yanchuk, Science (80-. ). \textbf{354}(6314), aag2472 (2016).
\newblock \doi{10.1126/science.aag2472}

\bibitem{Hsu2007a}
R.C.J. Hsu, A.~Ayazi, B.~Houshmand, B.~Jalali, Nat. Photonics \textbf{1}(9),
  535 (2007).
\newblock \doi{10.1038/nphoton.2007.145}

\bibitem{Sinha2015}
D.~Sinha, G.A.J. Amaratunga, Phys. Rev. Lett. \textbf{114}(14), 147701 (2015).
\newblock \doi{10.1103/PhysRevLett.114.147701}

\bibitem{Guha2006}
D.~Guha, Y.~Antar, IEEE Trans. Antennas Propag. \textbf{54}(9), 2657 (2006).
\newblock \doi{10.1109/TAP.2006.880766}

\bibitem{Ramo1994}
S.~Ramo, J.R. Whinnery, T.~{Van Duzer}, \emph{{Fields and waves in
  communication electronics}}, 3rd edn. (Wiley, New York, 1994)

\bibitem{DeYoung2006}
C.S. DeYoung, S.A. Long, IEEE Antennas Wirel. Propag. Lett. \textbf{5}(1), 426
  (2006).
\newblock \doi{10.1109/LAWP.2006.883952}

\bibitem{Guo2005}
Y.X. Guo, Y.F. Ruan, X.Q. Shi, IEEE Trans. Antennas Propag. \textbf{53}(10),
  3394 (2005).
\newblock \doi{10.1109/TAP.2005.856381}

\bibitem{Law2004}
C.J. Law, A.W. Roszak, J.~Southall, A.T. Gardiner, N.W. Isaacs, R.J. Cogdell,
  Mol. Membr. Biol. \textbf{21}(3), 183 (2004).
\newblock \doi{10.1080/09687680410001697224}

\bibitem{Alden1997}
R.G. Alden, E.~Johnson, V.~Nagarajan, W.W. Parson, C.J. Law, R.G. Cogdell, J.
  Phys. Chem. B \textbf{101}(23), 4667 (1997).
\newblock \doi{10.1021/jp970005r}

\bibitem{Novotny2007}
L.~Novotny, Phys. Rev. Lett. \textbf{98}(26), 266802 (2007).
\newblock \doi{10.1103/PhysRevLett.98.266802}

\bibitem{Yu2012}
Y.~Yu, V.E. Ferry, A.P. Alivisatos, L.~Cao, Nano Lett. \textbf{12}(7), 3674
  (2012).
\newblock \doi{10.1021/nl301435r}

\bibitem{Soren2014}
D.~Soren, R.~Ghatak, R.K. Mishra, D.R. Poddar, Prog. Electromagn. Res. B
  \textbf{60}(July), 195 (2014).
\newblock \doi{10.2528/PIERB14031306}

\bibitem{Alu2008a}
A.~Al{\`{u}}, N.~Engheta, Phys. Rev. B \textbf{78}(8), 085112 (2008).
\newblock \doi{10.1103/PhysRevB.78.085112}

\bibitem{Hjerrild2017}
N.E. Hjerrild, R.A. Taylor, Phys. Today \textbf{70}(12), 40 (2017).
\newblock \doi{10.1063/PT.3.3790}

\bibitem{Chaudhary2011}
R.K. Chaudhary, K.V. Srivastava, A.~Biswas, in \emph{2011 Natl. Conf. Commun.}
  (IEEE, 2011), pp. 1--5.
\newblock \doi{10.1109/NCC.2011.5734715}

\bibitem{Soren2012}
D.~Soren, R.~Ghatak, R.K. Mishra, D.R. Poddar, J. Electromagn. Anal. Appl.
  \textbf{04}(01), 9 (2012).
\newblock \doi{10.4236/jemaa.2012.41002}

\end{thebibliography}

\end{document}